\documentclass[%
 reprint,
 amsmath,amssymb,
 aps,%longbibliography
]{revtex4-2}

\usepackage{graphicx}% Include figure files
\usepackage{dcolumn}% Align table columns on decimal point
\usepackage{bm}
\usepackage{multirow}

\usepackage{mathtools}

\DeclarePairedDelimiter\ket{\lvert}{\rangle}

\usepackage{xcolor}
\newcommand{\cw}[1]{\textcolor{black}{#1}}
\newcommand{\cqw}[1]{\textcolor{black}{#1}}

\begin{document}

\title{Phase-controlled improvement of \cw{photon lifetime} in coupled superconducting cavities}

\author{Changqing Wang}\email{cqwang@fnal.gov}
\author{Oleksandr S Melnychuk}
\author{Crispin Contreras-Martinez}
\author{Yao Lu}
\author{Yuriy M Pischalnikov}
\author{Oleg Pronitchev}
\author{Bianca Giaccone}
\author{Roman Pilipenko}
\author{Silvia Zorzetti}
\author{Sam Posen}
\author{Alexander Romanenko}\email{aroman@fnal.gov}
\author{Anna Grassellino}\email{annag@fnal.gov}
\affiliation{Superconducting Quantum Materials and Systems Center, Fermi National Accelerator Laboratory (FNAL), Batavia, IL 60510, USA}
\date{\today}

\begin{abstract}
\cw{High-quality} cavities are crucial for various fundamental physical studies and applications. Here we find that by coupling two cavities directly or via a phase-tunable coupling channel, the \cw{photon lifetime} of the local field can exceed that of the bare cavities. The cavity \cw{photon lifetime} is modified by the phases of the initial states and the phase accumulation on the coupling channel which affect the interference between cavities. In experiments, by coupling superconducting radio-frequency cavities via phase-tunable cables, we realize a factor of two improvement in the cavity \cw{photon lifetime}. The results can bring rich revenue to quantum information science, sensing, and high-energy physics.
\end{abstract}

\maketitle

\makeatletter
\def\l@subsubsection#1#2{}
\makeatother

\section{Introduction}
High-quality (high-Q) cavities, the resonant structures that preserve electromagnetic energy with low dissipation rates, are essential devices in quantum sciences, high-energy physics, and modern industries. Particularly, in the recently fast-growing quantum information sciences, \cw{high-quality} superconducting radio-frequency (SRF) cavities \cite{padamsee2014superconducting,romanenko2014ultra} play a pivotal role in constructing quantum memory, quantum processing units (QPUs), and sensors \cite{alam2022quantum,berlin2022searches,reagor2016quantum, paik2011observation}. In those devices, the \cw{photon lifetime (or energy decay time)} of a cavity, which is the \cw{duration for intracavity photons to undergo energy decay subject to dissipation effects}, \cqw{significantly affects} the performance of quantum devices, such as gate operating time in QPUs, sensitivity of sensors, etc. \cqw{Moreover, in systems utilizing cavities for qubit readout, there is a trade-off between the readout speed and the cavity-induced decoherence due to the Purcell effect.} Therefore, the capability to engineer cavity \cw{photon lifetime} is crucial for achieving high performance in realistic physical devices and systems. 

Past decades have witnessed great progresses in mitigating dissipation and decoherence in microwave/optical cavities with state-of-the-art material and fabrication techniques \cite{romanenko2017understanding}. Two-second photon lifetime has been reported in three-dimensional (3D) niobium SRF cavities at the quantum regime \cite{romanenko2020three}. However, \cqw{the cavity photon lifetime in realistic quantum architectures is often constrained by various factors such as the inverse Purcell effect and dielectric losses in circuit quantum electrodynamics \cite{wang2008measurement, koch2007charge, blais2021circuit} and hybrid quantum systems \cite{han2021microwave,lauk2020perspectives, zorzetti2023milli,wang2022high,wollack2021loss,goryachev2015single}}. Therefore, further approaches are demanded to \cqw{control photon lifetime and energy relaxation processes} in cavities. 

\cw{While a single cavity provides a simple platform for many problems, richer dynamic behavior can be found in} systems of two or more cavities coupled directly or indirectly. Past studies have shown unconventional phenomena in coupled-mode systems, including \cw{quantum state transfer \cite{axline2018demand}, qubit state encoding and error correction \cite{chou2023demonstrating}}, single-photon state generation \cite{majumdar2012loss}, high-precision sensing \cite{chen2017exceptional,zhang2019quantum}, absorption control \cite{wang2020electromagnetically, wang2021coherent}, etc. It has also been proposed that coupled-mode systems exhibit modified decoherence processes under non-Hermitian effects \cite{gardas2016pt,dey2019controlling,partanen2019exceptional}. Moreover, it has been shown that the phase of coupling parameters in a coupled cavity system poses nontrivial influences on quantum statistical properties of bosonic states \cite{wang2017phase}. Such phase effect on the field dynamics and photon lifetime in coupled cavity systems has not been fully investigated.

In this letter, we show that the phase-controlled interference affects the photon lifetime of the local field in coupled-cavity systems. We experimentally demonstrate a slow-down of photon number decay in two SRF cavities coupled through a coaxial cable with phase control at cryogenic temperatures. 

\section{Theoretical model of the coupled cavity systems}
We study a system where two near-resonant cavities are coupled to each other, with two types of coupling considered: direct coupling via field overlap and indirect coupling via an intermediate mode in a cable or waveguide. 

\begin{figure}[!htb] 	\centering\includegraphics[width=0.46\textwidth]{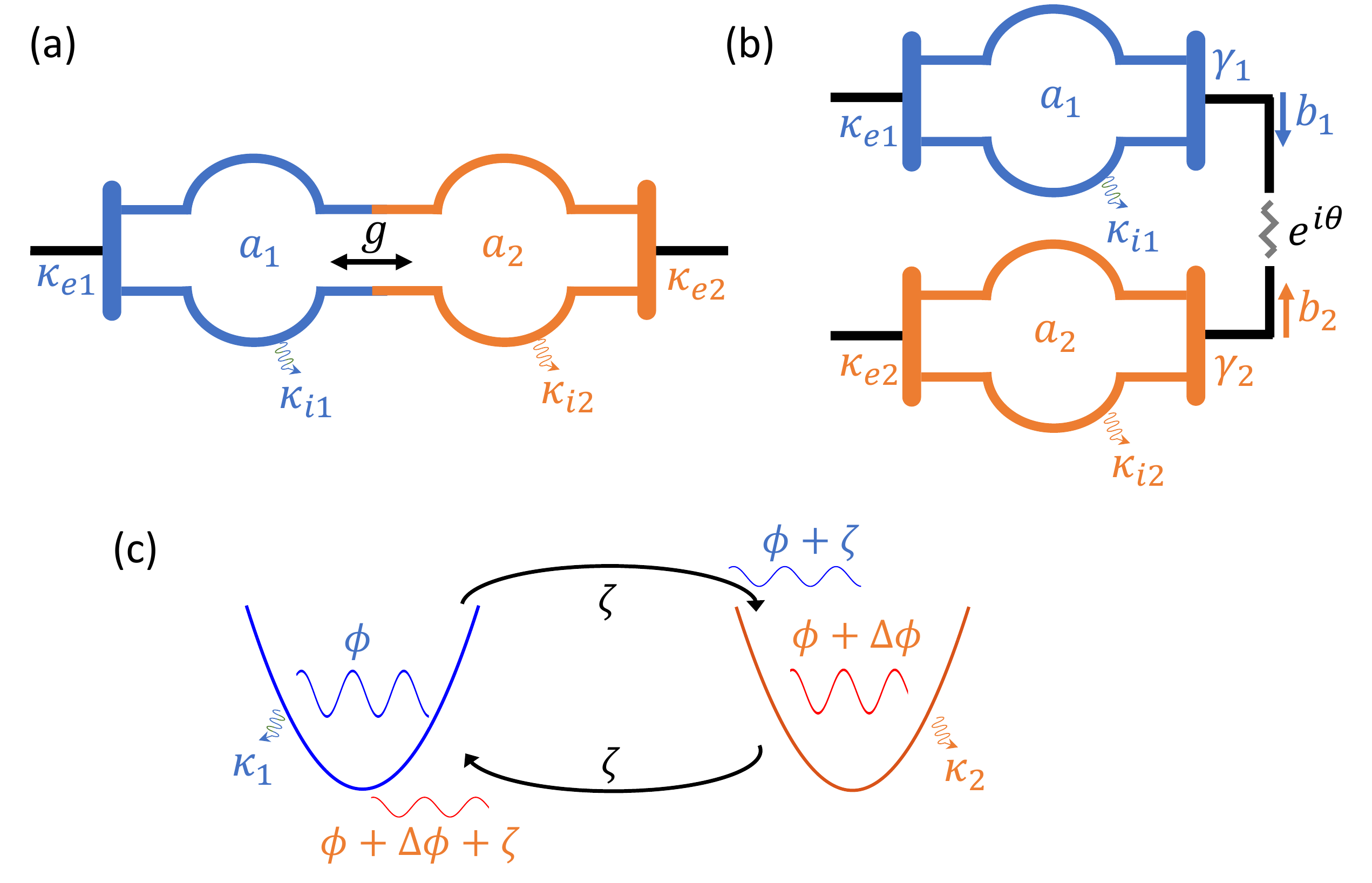}
	\caption{Interference effects in coupled-cavity systems. (a) Schematic diagram of two SRF cavities with internal loss rates $\kappa_{i1,i2}$ directly coupled with coupling strength $g$. The cavities are coupled to the left and right ports with the coupling strength $\kappa_{e1,e2}$. (b) Schematic diagram of two SRF cavities coupled to an intermediate cable/waveguide with coupling strength $\gamma_{1,2}$. The propagation on the cable/waveguide yields $\theta$ phase shift. The definitions of other parameters are the same as in (a). (c) Illustration of field interference within the coupled cavities. The intracavity fields (sinusoidal curves with large amplitudes) have initial phases of $\phi$ and $\phi+\Delta\phi$, respectively. The fields tunneling from one cavity to another (sinusoidal curves with smaller amplitudes) are accompanied by a \cw{$\zeta$} phase shift.
}
	\label{Fig1}
\end{figure}

The direct coupling can be realized via the overlap of cavity mode profiles [Fig.~\ref{Fig1}(a)]. Assuming the cavities support modes $a_{1,2}$ with frequencies $\omega_{1,2}$, and are coupled with the coupling strength $g$, we can write the system Hamiltonian as
\begin{equation}
    H=\hbar\omega_1a_1^\dag a_1+\hbar\omega_2a_2^\dag a_2+g(a_1^\dag a_2+a_1a_2^\dag) .\label{H}
\end{equation}
\cw{The full quantum dynamics of the system can be described by the Lindblad master equation.  One can neglect the quantum jump terms in several senarios: 1) by operating within the semiclassical limit \cite{roccati2022non}, 2) through the post-selection of states to eliminate quantum trajectories with quantum jumps \cite{naghiloo2019quantum}, or 3) in cases of extremely weak excitation where the cavity state is close to the ground state \cite{wang2023non,wang2017phase}.} The dynamic equations for the mode operators are  
\begin{align}
\frac{d}{dt}
\begin{bmatrix}
a_1 \\
a_2 
\end{bmatrix}
    &=
\begin{bmatrix}
-i\omega_1-\frac{\kappa_1}{2} & -ig\\
-ig & -i\omega_2-\frac{\kappa_2}{2} 
\end{bmatrix}
\begin{bmatrix}
a_1 \\
a_2 
\end{bmatrix}\nonumber
\\&-
\begin{bmatrix}
\sqrt{\kappa_{e1}}a_{in,1} \\
\sqrt{\kappa_{e2}}a_{in,2} 
\end{bmatrix},
\label{L1}
\end{align}
where $a_{in,1,2}$ are the input fields. The total loss rates $\kappa_{1,2}$ are composed of the internal and external losses, i.e., $\kappa_{1,2}=\kappa_{i1,i2}+\kappa_{e1,e2}$.

The indirect coupling is realized via an intermediate coaxial cable/waveguide that is coupled to both cavities with the coupling strengths $\gamma_{1,2}$ [Fig.~\ref{Fig1}(b)]. The wave propagation along the cable/waveguide with length $L_0$ leads to a $\theta$ phase shift and additional loss described by a linear coefficient $\gamma_0$. $b_{1}$ ($b_{2}$) is the field coupled from $a_{1}$ ($a_{2}$) to the left (right) end of the cable/waveguide. The propagation on the cable/waveguide picks up a prefactor $e^{i\theta-\gamma_0L_0/2}$ due to phase accumulation and dissipation. \cw{Using the input-output relation \cite{combes2017slh}} yields the dynamic equations
\begin{align}
    &\frac{da_1}{dt}=(-i\omega_1-\frac{\kappa_1}{2})a_1-\sqrt{\gamma_1} b_2e^{i\theta-\gamma_0L_0/2}-\sqrt{\kappa_{e1}}a_{in,1},\nonumber\\
    &\frac{da_2}{dt}=(-i\omega_2-\frac{\kappa_2}{2}) a_2-\sqrt{\gamma_2} b_1e^{i\theta-\gamma_0L_0/2}-\sqrt{\kappa_{e2}}a_{in,2},\nonumber\\
    &\cw{b_1=b_2e^{i\theta-\gamma_0L_0/2}+\sqrt{\gamma_1}a_1,}\nonumber\\
    &\cw{b_2=b_1e^{i\theta-\gamma_0L_0/2}+\sqrt{\gamma_2}a_2,}
\label{L2}
\end{align}
where $\kappa_{1,2}=\kappa_{i1,i2}+\kappa_{e1,e2}+\gamma_{1,2}$. By eliminating $b_{1,2}$, one can obtain

\cw{
\begin{align}
\frac{d}{dt}
\begin{bmatrix}
a_1 \\
a_2 
\end{bmatrix}
    &=
\begin{bmatrix}
-i\omega_1-\frac{\kappa_1}{2}-i\delta \omega_1 & -ig_{eff} \\ -ig_{eff} & -i\omega_2-\frac{\kappa_2}{2}-i\delta\omega_2 
\end{bmatrix}
\begin{bmatrix}
a_1 \\
a_2 
\end{bmatrix}\nonumber
\\&-
\begin{bmatrix}
\sqrt{\kappa_{e1}}a_{in,1} \\
\sqrt{\kappa_{e2}}a_{in,2} 
\end{bmatrix},\label{L3}
\end{align}
where $\delta\omega_{1,2}=-i\frac{\gamma_{1,2}e^{2i\theta-\gamma_0L_0}}{1-e^{2i\theta-\gamma_0L_0}}$, and $g_{eff}=-i\frac{\sqrt{\gamma_1\gamma_2}e^{i\theta-\gamma_0 L_0/2}}{1- e^{2i\theta-\gamma_0L_0}}$.}

\cqw{In the rotating frame $a_{1,2}=A_{1,2}e^{-i\omega t}$ with an initial condition $A_{1,2} (t=0)=A_{10,20}$, one can obtain 
\begin{align}
A_{1,2} (t)&=A_{10,20} e^{-i\Delta_{1,2} t-\frac{\kappa_{1,2eff}}{2} t}+ A_{20,10} e^{-ig_{eff} t},
\end{align}
where $\Delta_{1,2}=\omega-\omega_{1,2}-Re(\delta\omega_{1,2})$, and $\kappa_{1,2eff}=\kappa_{1,2}-2Im(\delta\omega_{1,2})$.} One notable effect is that the interference of two cavity modes can induce distinct dynamics of the local field, leading to prolonged or shortened \cw{photon lifetime}, as discussed below. 

\section{Interference effects in coupled cavities}

We illustrate the field interference process within indirectly coupled cavities in Fig.~\ref{Fig1}(c). We assume that the cavity fields have initial phases of $\phi$ and $\phi+\Delta\phi$, respectively. According to Eq.~(\ref{L3}), the tunneling of the field from one cavity to another via the indirect coupling scheme leads to a phase shift \cw{$\zeta=3\pi/2+arg(g_{eff})$}, where $3\pi/2$ originates from the $-i$ factor in the anti-diagonal term of the dynamic matrix. The same applies to the field tunnelling from cavity 2 to cavity 1. \cw{It is noted that as the cable/waveguide loss is small, $\zeta$ is approximately $3\pi/2$ which corresponds to an energy conservation case.}

As a result, interference occurs between the field tunneling from cavity 1 to cavity 2 with a phase $\phi+\zeta$ and the existing field in cavity 2 with a phase $\phi+\Delta\phi$. The interference is constructive (destructive) if these two fields are in phase (completely out of phase), with the conditions summarized in Table~\ref{table1}.

\begin{table}[b]
\centering
\caption{Conditions of constructive and destructive interference for indirectly coupled cavities. $\Delta\phi$ represents the initial phase difference between the fields in cavity 2 and cavity 1. $\zeta$ denotes the phase accumulation along the intermediate cable/waveguide. $m$ represents an arbitrary integer.}\label{table1}
\begin{ruledtabular}
\begin{tabular}{ccc}
Interference type & Condition\\
\hline
 Constructive interference in cavity 1 & $\Delta\phi+\zeta=2m\pi$ \\ 
Destructive interference in cavity 1 & $\Delta\phi+\zeta=(2m+1)\pi$ \\
Constructive interference in cavity 2 & $\Delta\phi-\zeta=2m\pi$ \\ 
 Destructive interference in cavity 2 & $\Delta\phi-\zeta=(2m+1)\pi$ \\ 
\end{tabular}
\end{ruledtabular}
\end{table}

%As for the indirect coupling scheme with a lossy channel, $g_{eff}=\frac{\sqrt{\gamma_1\gamma_2}}{\sin{(\theta+i\gamma_0L_0/2)}}=\frac{\sqrt{\gamma_1\gamma_2}}{\sin(\theta)\cosh(\gamma_0L_0/2)+i\cos(\theta)\sinh(\gamma_0L_0/2)}$, and the effective phase shift for field to tunnel across the cavities becomes $3\pi/2-\arctan[\cot(\theta)\tanh(\gamma_0L_0/2)]$. 

%For a lossless cavity link, $\gamma_0=0$ leads to $g_{eff}=\frac{\sqrt{\gamma_1\gamma_2}}{\sin{(\theta)}}$)

The same physical mechanism can be applied to the directly coupled cavities, where the total phase shift for the field tunneling from one cavity to another is \cw{$\zeta=3\pi/2$}. Consequently, constructive interference in cavity 1 is always associated with destructive interference in cavity 2, and vice versa. The conditions for constructive and destructive interference are summarized in Table~\ref{table2}.

\begin{table}[b]
\centering
\caption{Conditions of constructive and destructive interference in directly coupled cavities. The definitions of parameters are the same as Table~\ref{table1}.}\label{table2}
\begin{ruledtabular}
\begin{tabular}{ccc}
Cavity 1 & Cavity 2 & Condition\\
\hline
Constructive & Destructive & $\Delta\phi=(2m+\frac{1}{2})\pi$ \\ 
Destructive & Constructive & $\Delta\phi=(2m-\frac{1}{2})\pi$ \\
\end{tabular}
\end{ruledtabular}
\end{table}

\section{The phase-controlled interference effect on cavity photon lifetime}
The \cw{photon lifetime} of local fields in coupled-cavity systems can be modified by the interference effect. Intuitively, the constructive interference compensates the decay of the intracavity energy, while the destructive interference enhances the energy dissipation. \cw{In addition, Eq.~(\ref{L3}) shows that the indirect coupling leads to complex frequency shifts $\delta\omega_{1,2}$ in both cavity modes, with the real part representing a frequency shift and the imaginary part a modification of the loss rates, both depending on $\theta$. Therefore the optimized $\theta$ value may be shifted from the conditions listed in Table.~\ref{table1}.} %For simplicity, in our consideration where the losses on the cable/waveguide is relatively small, we can choose $\theta$ such that the induced loss is minimal. When cable/waveguide losses are large, the induced losses could also reduce the interference effect.

\begin{figure}[!htb] 
	\centering
	\includegraphics[width=0.46\textwidth]{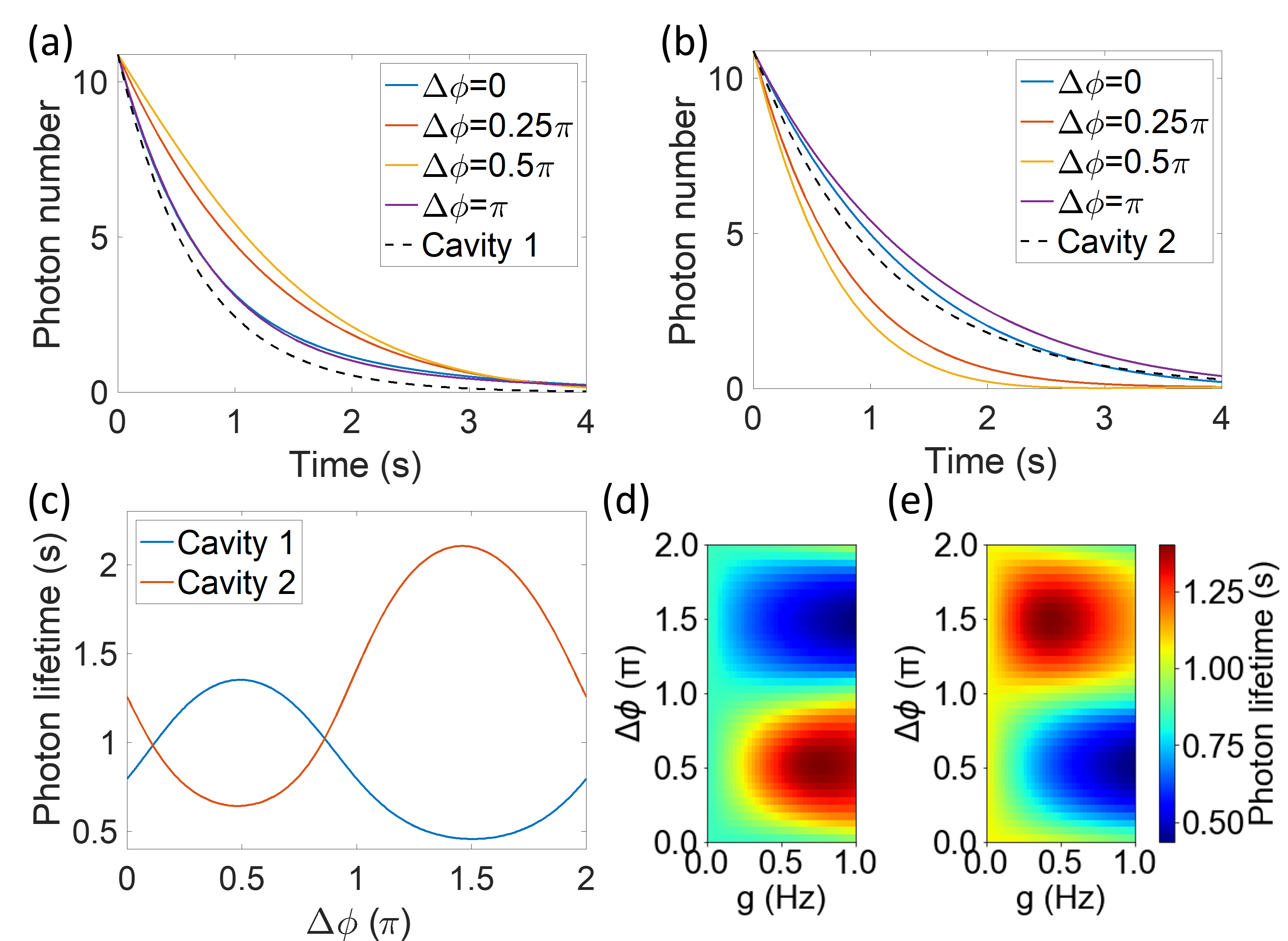}
	\caption{\normalsize Simulation of the state evolution in coupled-cavity systems. (a)-(c) State evolution in indirectly coupled cavities, based on Eq.~(\ref{L3}). \cw{$\theta=\pi/10$, $\gamma_0L_0=0.02$, yielding $\zeta=(1.5-0.0098)\pi$}. (a)(b) Time evolution of intracavity photon numbers in cavity 1 (a) and cavity 2 (b) is plotted for various initial phase difference $\Delta\phi$. (c) \cw{Photon lifetime} of cavity 1 and cavity 2 versus $\Delta\phi$. (d)(e) Time evolution of directly coupled cavities solved by Python QuTip. \cw{Photon lifetime} of cavity 1 (d) and cavity 2 (e) are plotted as a function of the initial phase difference $\Delta\phi$ and the coupling strength $g$ between the two cavities. Cavity temperature is 10mK for (d) and (e).
}
	\label{Fig2}
\end{figure}

To determine the relation between the phases, interference, and \cw{photon lifetime}, we investigate the cavity field evolution with the initial condition that two cavitites are at coherent states $\ket{\alpha}$ and $\ket{\alpha e^{i\Delta\phi}}$, respectively, with a phase difference ($\Delta\phi$) in their amplitudes. By solving Eq.~(\ref{L3}), we find that the decay rates of cavity photon numbers depend on $\Delta\phi$ [Fig.~\ref{Fig2}(a)(b)], and vary from that of an uncoupled cavity. 

We \cw{characterize the photon lifetime by measuring} the time for cavity energy to decay to $1/e$ of its initial value. Figure~\ref{Fig2}(c) shows the variation of \cw{photon lifetime} for both cavities against $\Delta\phi$. The \cw{photon lifetime} is optimized around $\Delta\phi=1.5\pi$ for $\theta=\pi/10$, \cw{as $\zeta$ is approximately $1.5\pi$}, which matches the constructive interference condition in Table.~\ref{table1}.  

The prolonged \cw{photon lifetime} is also observed in directly coupled cavities. Python QuTip toolbox is used to simulate the evolution of few-photon coherent states governed by Lindblad master equations in two directly coupled cavities  
                    prepared at coherent states $\ket{1}$ and $\ket{e^{i\Delta\phi}}$, respectively. Figures~\ref{Fig2}(d)(e) show that the \cw{photon lifetime} of both cavities oscillates with $\Delta\phi$, and reaches maximum at $g=$~0.3Hz. Furthermore, at $\Delta\phi=\pi/2$, the longest (shortest) \cw{photon lifetime} is observed for cavity 1 (2) and vice versa for $\Delta\phi=3\pi/2$, which are consistent with the conditions in Table.~\ref{table2}. In both the direct and indirect coupling cases, the \cw{photon lifetime} can exceed that of each uncoupled cavity. This provides a phase-dependent approach to mediate cavity dissipation without changing the cavity material or structural property. 

Such effect observed in the transient dynamics depends on the initial state of the system. The initial state corresponding to the longest \cw{energy decay time} in one cavity is in neither eigenmode of the system, but rather a superposition of two. The interference between the two eigenmodes with a frequency difference ($\Delta f=\sqrt{4 g^2-({\kappa_1-\kappa_2})^2/4}$ for $\omega_1=\omega_2$), lead to a slow oscillation that modifies the decay profile, though the Q factors of eigenmodes do not outperform that of the individual cavity. 

\section{Experimental observation of photon lifetime improvement}
While the preparation of arbitrary initial states is nontrivial, one feasible case is to prepare steady states by continuous one-port excitation. Assuming $\omega_1=\omega_2$ and zero input detuning in Eq.~(\ref{L3}), the steady-state fields in two cavities have a phase difference $\Delta\phi=\theta+\pi$. This satisfies the constructive interference condition in cavity 2 once the input is off, leading to \cw{a favorable regime to improve the photon lifetime} in cavity 2. 

\begin{figure}[!htb] 
	\centering
	\includegraphics[width=0.46\textwidth]{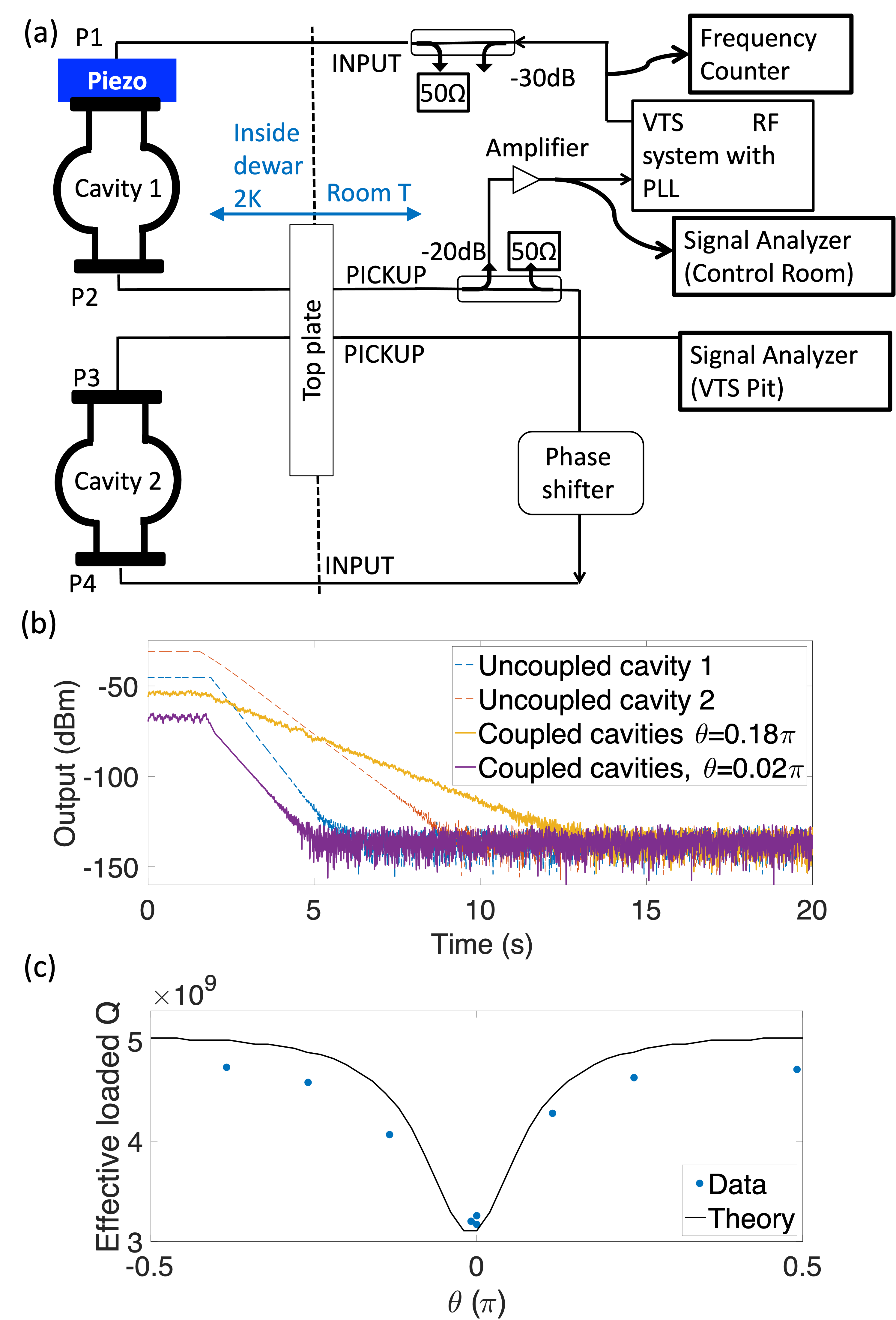}
	\caption{\normalsize Experimental study of the \cw{photon lifetime} of indirectly coupled cavities. (a) Schematic diagram of the experimental setup based on the vertical test system (VTS) \cqw{where cavities are immersed into liquid helium in a vertical dewar at 2~K}. The RF signal is injected to the P1 port of cavity 1. 1\% of the output from P2 is picked up as the feedback signal for the phase lock loop (PLL), while the remaining signal passes a phase shifter and reaches cavity 2. The total transmission spectrum and the time-domain output power is measured from P3 port using a signal analyzer. (b) Measured transmitted power decay for coupled cavities at 1.3~GHz, with different $\theta$, compared with uncoupled cavity 1 and 2. (c) Effective Q factor vs $\theta$ extracted from the power decay of a pair of 2.6~GHz SRF cavities \cqw{measured via a spectrum analyzer}. 
}
	\label{Fig3}
\end{figure}

We experimentally demonstrate such effects in a system composed of two \cqw{niobium} TESLA SRF cavities \cite{aune2000superconducting} with similar frequencies in their fundamental modes aligned by the piezo actuator which induces a mechanical displacement on the wall of cavity 1 [Fig.~\ref{Fig3}(a)]. \cqw{High Q factors were achieved via heat treatment in the custom designed oven to remove the niobium pentoxide \cite{romanenko2020three, berlin2022searches}}. At the four coupling ports (P1-P4) antennas are mounted on the flanges to couple the coaxial cables with the cavities. A tunable phase shifter is inserted to the cable between cavities for phase control. 

We first choose two cavities with 1.3~GHz resonance frequencies and \cw{loaded $Q$ factors $Q_{L1}=1.53\times10^{9}$, $Q_{L2}=2.56\times10^9$,} respectively. The transmission spectrum collected from P2 allow the monitoring of two resonance peaks. After matching the resonance frequencies with the piezo actuator, we lock the signal to the resonance and perform decay measurement by turning off the input and monitoring the output power \cqw{via a spectrum analyzer}. The measurement is performed with different values of $\theta$ [Fig.~\ref{Fig3}(b)]. At $\theta=0.18\pi$, we observe an extended \cw{photon lifetime} from cavity 2, corresponding to an effective Q factor ($4.98\times10^9$) nearly twice as large as the maximum loaded $Q$ for uncoupled cavity 1 and 2 ($2.56\times10^9$). Furthermore, $\theta=0.02\pi$ leads to shorter \cw{photon lifetime} than that of cavity 2. Such a variation attributes to the phase-dependent effective loss on the coupling channel. 

The coherence enhancement also appears for the case that $Q_{L1} \ll Q_{L2}$. We use two 2.6~GHz cavities with \cw{$Q_{L1}=4.29\times10^8$ and $Q_{02}=4.46\times10^9$}, respectively. The measured effective loaded Q varies with $\theta$ [Fig.~\ref{Fig3}(c)]. In a large range of $\theta$ the effective $Q$ shows advantage over $max(Q_{L1},Q_{L2})=4.46\times10^9$, indicating that the \cw{photon lifetime} can be prolonged by coupling to a cavity with a much lower $Q$. 

\begin{figure}[!htb] \centering\includegraphics[width=0.46\textwidth]{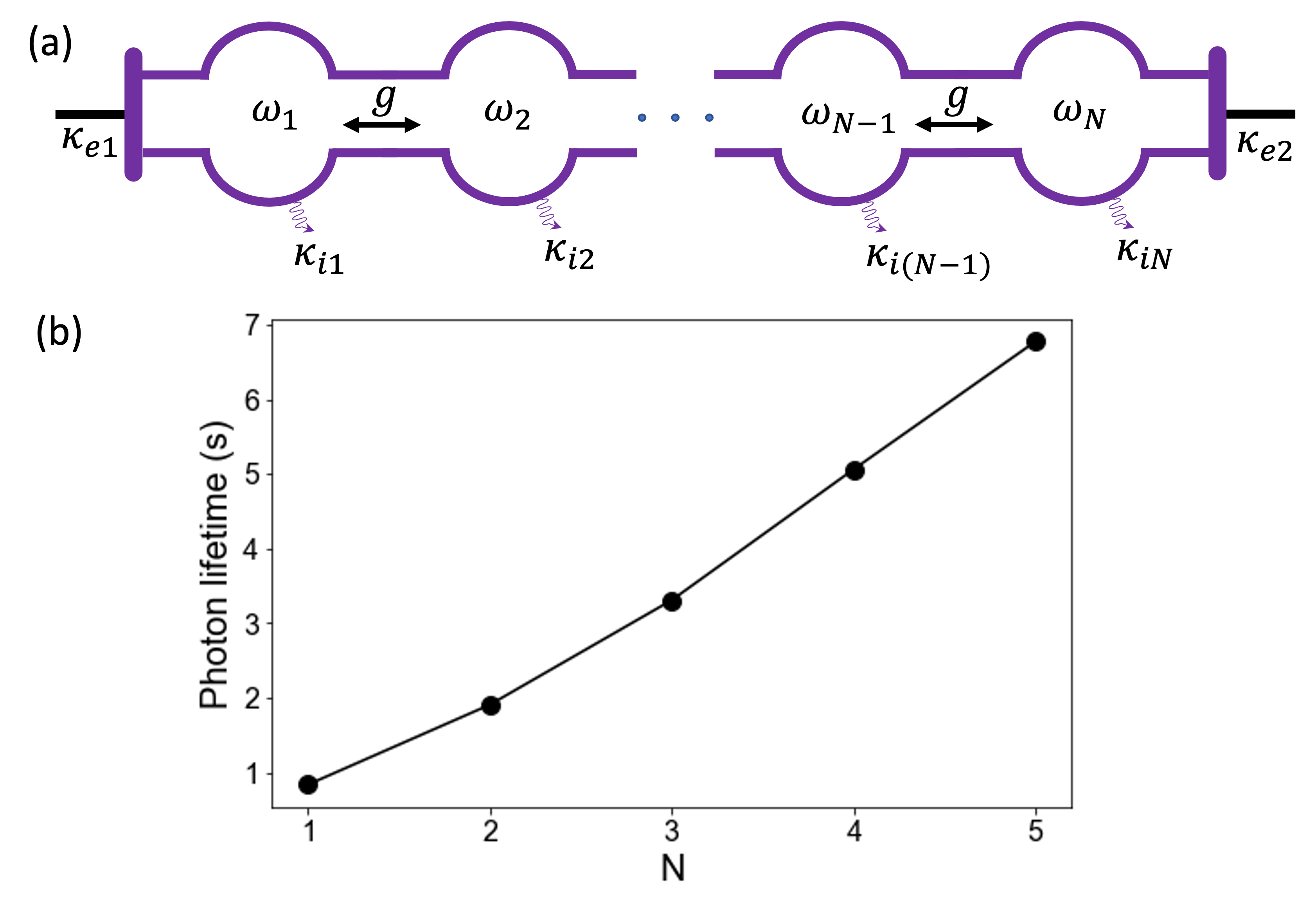}
	\caption{\normalsize \cw{Photon lifetime improvement} in a N-cavity system. (a) Schematic diagram of N cascaded coupled SRF cavities. The first and Nth cavities are coupled to antennas with the coupling strength $\kappa_{1,2,ext}$, respectively. The adjacent cavities are coupled with the coupling strength $g$. (b) \cw{Photon lifetime} of the $N$th cavity as a function of $N$. Parameters: $g=$0.2~Hz, $\kappa_{i,j}=$1~Hz ($j=1,2,...,N$), $\kappa_{e1}=\kappa_{e2}=$0.2~Hz.
}
	\label{Fig4}
\end{figure}

\section{Extension to N coupled cavities}
We further extend our discussions to a cascaded system composed of N directly coupled cavities [Fig.~\ref{Fig4}(a)]. For left-port excitation, the phase difference between two adjacent cavities is $-\pi/2$ at steady states, leading to constructive interference and field compensation from left to right on the entire cavity chain. Simulation results show that the \cw{photon lifetime} of the Nth cavity measured from the right port, scales approximately linearly with the total number of cavities [Fig.~\ref{Fig4}(b)]. This indicates a large room of lifetime enhancement and readout speed control in such a cascaded structure. 

\section{Conclusion and outlook}
In directly and indirectly coupled cavities, we find considerable improvement in \cw{photon lifetime} of local fields by properly engineering the initial phase of the cavity states and phase accumulation on the coupling channel. \cw{The observed improvement of photon lifetime provides a pathway to mitigate the energy losses in the local intracavity field without any material or fabrication modification, thereby enhancing the performance of cavity-based devices.} \cqw{While the intrinsic quality factor constrains the quantum coherence in many cases, the modified cavity photon lifetime can still significantly affect coherence time of qubits via the Purcell effect. As the phase control provides a tunable speed of energy decay from a cavity, such a system coupled to a qubit can act as a Purcell filter to manipulate the qubit readout time with phase control without adding additional Purcell loss. Alternatively, one can achieve fast qubit readout through phase control using cavities with higher intrinsic Q factors, which leads to further protection of qubit coherence time.} Such a scheme can \cqw{enhance the performance of various systems including} superconducting QPUs, optical/mechanical/acoustic devices, and hybrid quantum systems.

\begin{acknowledgments}
This material is based upon work supported by the U.S. Department of Energy, Office of Science, National Quantum Information Science Research Centers, Superconducting Quantum Materials and Systems Center
(SQMS) under contract number DE-AC02-07CH11359.
\end{acknowledgments}

\bibliography{references}

\end{document}